\documentclass{eas}
\usepackage{graphicx}
\usepackage{natbib}
\usepackage{amssymb}
\begin{document}

\title{Diffusion and pulsations in slowly rotating B stars}

\runningtitle{Turcotte \& Richard: Diffusion and pulsations in B stars \dots}
\author{Sylvain Turcotte}
\address{Physics Dept., Bishop's University, Lennoxville, Qc, Canada, J1M 1Z7}
\author{Olivier Richard}
\address{GRAAL, Universit\'e Montpellier II, cc072, Place Eug\`ene Bataillon, 34095 Montpellier, Cedex 5, France}
%
%
\begin{abstract}
Diffusion in cool B stars of the main sequence has been shown to strongly 
affect opacities and convection in cool B stars of the main sequence. 
We show here that diffusion in B stars maintains or enhances the excitation of
pulsations in these stars. This result conflicts with observations as cool B stars
that show evidence of diffusion, the HgMn stars, are stable to the current
detection level. 
We discuss possible implications of this discrepancy for the models.
\end{abstract}
\maketitle

\section{Diffusion and opacities in cool B stars}

Slowly rotating B stars on the main sequence are thought to be extremely stable as
there is only limited convection in the outer envelope and none at the surface, mass
loss is expected to be small and rotational mixing is also negligible.
In such a stable environment, diffusion should proceed with few impediments.
Consequently, the large abundance anomalies observed in HgMn stars are understood to be 
the result of diffusion in the atmosphere of such stars.


That diffusion occurs in the atmosphere suggests that diffusion also occurs in the interior
in the absence of mixing processes. 
Richer, Michaud \&~Turcotte~(2000) and Richard, Michaud \&~Richer~(2002) (see also Richard 
in these proceedings) have shown that
radiative levitation pushes iron-peak elements up in the envelope of hot A and 
cool B stars. While these elements are radiatively supported throughout
the outer envelope of these stars, they tend to accumulate at a temperature of 
roughly 200\,000~K because of a local reduction of the outward flux there.
At such a temperature, iron-peak elements are the dominant contributors to the opacity 
and, naturally, as they accumulate, the opacity also increases locally. 
Figure~1 shows the abundance profiles of two models, one with significant diffusion and
another where the effect of diffusion is only marginal. An overabundance of the order 
of a factor of ten is achieved in the former.

With the notable exception of Hydrogen, lighter elements that play an important 
role in the opacity at lower temperature 
are generally not supported by radiative pressure and therefore sink out of the superficial
regions toward the core. Consequently, their contribution to 
the opacity diminishes.

The combined effect of the evolution of the chemical composition as a result of 
diffusion yields a marked increase
in the opacity at around 200\,000~K and, perhaps somewhat surprisingly, an increase
of the opacity at lower temperatures, as shown in Figure~1, due to the combined increase
in the opacity due to hydrogen and iron. 

Notice that the abundances are homogeneous from the surface to a point deeper than
200\,000~K ($\log T =5.3$). These low temperature regions are artificially homogenized
with an ad hoc turbulent mixing coefficient. In one respect this allows us to tweak the
level of chemical anomaly in order to investigate the effect of diffusion in the star's
interior. Unfortunately, it also means that the structure of the cool regions of
the envelope may be significantly inaccurate.

\section{Pulsations in cool B stars}

The kappa-mechanism due to the opacity of iron-peak elements
is responsible for variability on the lower main sequence (Pamyatnykh 1999). 
The SPB stars (Slowly Pulsating B stars; see Pamyatnykh) are long-period pulsators
found in chemically normal young main-sequence stars earlier than B8. 
The distribution of these stars overlap in the H.-R. diagram those of
the chemically peculiar but seemingly stable HgMn stars.

Apart from variability and surface chemical composition these two classes of
stars are very similar. Interestingly, SPB stars are found to be mostly slowly rotating
stars,  as are the HgMn stars, but the lack of rapidly rotating SPB stars may well be
only a selection effect.
This suggests that there might well be a correlation between the chemical composition 
and the excitation of the pulsations, as in Am stars where diffusion leads to
stability, or that the conditions that allow diffusion are
not conducive to pulsations occurring.

The best models currently available (Turcotte \& Richard, submitted) do not however 
support the hypothesis that diffusion can undermine the excitation of pulsations.  
As the opacity bump due to iron-peak is enhanced as a result of diffusion
in those models, they suggest that the excitation of pulsations in HgMn stars should
be at least as high than in chemically normal SPB stars. Again in Figure~1, the differential work 
for a given mode of pulsation is shown in a model with nearly normal composition 
and one with strongly enhanced
iron and opacity in the driving region. The net normalized growth rate, which must be positive
for a mode to be unstable and for the star to become variable, for this
mode is 0.08 in the ``normal'' model and 0.22 in the ``peculiar'' model. The 
peak in the driving region is higher, but there is also more damping on the 
hot side of the peak.  In this mode the net driving is in fact considerably enhanced,
but in many modes, especially in more evolved models, the net excitation (the value 
of the normalized growth rate) is surprisingly insensitive to the 
magnitude of the abundance anomalies.
Nevertheless, one must conclude that the models are lacking the necessary ingredient 
to explain the lack of observed pulsations in HgMn stars.
\begin{figure}
  \includegraphics[width=13cm]{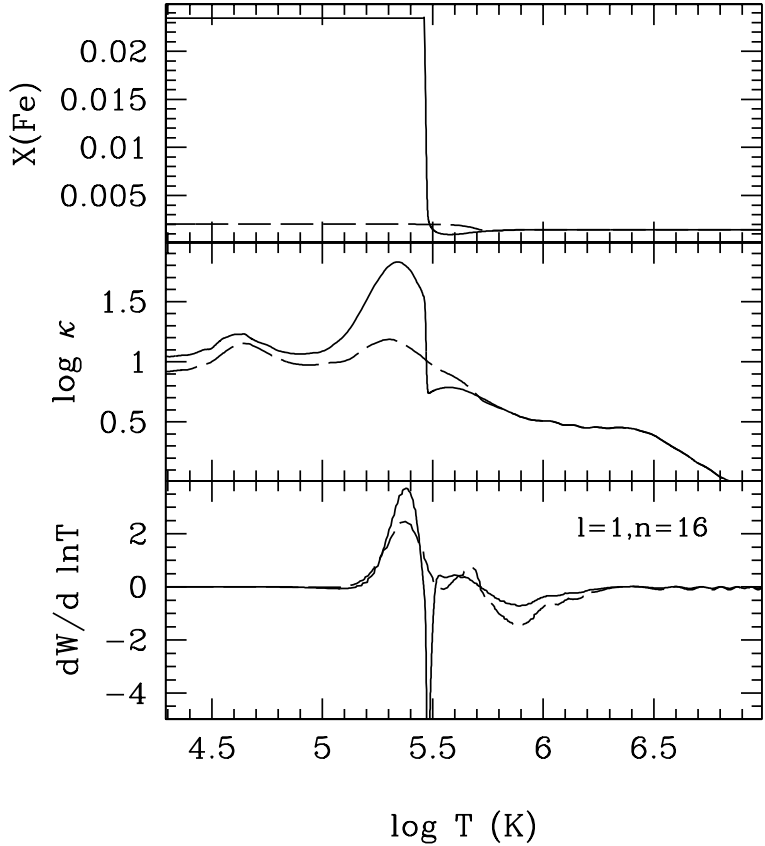}
\caption{The figure illustrates the effect of diffusion on the excitation of pulsations
in the model of a 10~Myr old 4~M$_\odot$ star. Two models of the same age and mass are compared,
one with only marginal change in abundances (dashed line) and one with more efficient
diffusion (solid line). The top panel shows the iron abundance profiles;
the middle panel the mean Rosseland opacity; and the bottom panel shows the 
differential work for a $\ell=1, n=16$ mode with a period of 1.2 days. A positive work indicates
mode excitation while a negative value indicates damping. By integrating the differential work
over the whole star we obtain the total work integral which is used to calculate the 
normalized growth rate (see text).}
\end{figure}

\section{What does this tell us about cool B stars?}

We can speculate as to what is the missing ingredient in the models.
 
We can first argue that adding mixing in the interior of HgMn stars 
would not resolve the discrepancy as the result of mixing would be to homogenize
the composition to its initial, here solar, value. This would still leave too much iron-peak
elements in the pulsations' driving region, leading to the expectation
of pulsations in HgMn stars as in SPB stars.

A possible solution may be the selective mass loss of certain elements from radiation
pressure in the atmosphere but not others (Babel 1995).  It is possible that this
may lead to the depletion of some elements in the driving region. 
The detailed process by which this depletion would occur, if indeed it can,
has not been worked out yet. 

Another possibility is that the issues of mode selection, interference or
visibility that often befall the asteroseismology of pulsating stars obscures
any direct conclusions we can hope to make on models and stellar physics 
from the observations. 

Finally, our models are lacking in one crucial aspect.
Our current models cannot model the region cooler than 200\,000~K consistently 
because of numerical problems. Therefore the structure of the models there may not
be appropriate. Though the work integrals seem rather insensitive to those regions,
a substantial change in structure there may lead to smaller predicted excitations.

The major stumbling block to improved models is the lack of
opacity spectra appropriate to model diffusion consistently at low temperatures 
(Leblanc, Michaud \&~Richer~2000).  Only when this will be possible will the 
full picture of mode driving in HgMn stars be achieved. Before then, the models 
remain informative of the processes that occur in the interior, but speculative
as to the net effect of diffusion of mode damping.

Observationally, the advent of space-based experiments dedicated to asteroseismology
will eventually resolve the question of whether HgMn stars are really stable or if 
they undergo undetected low-amplitude variations. Observations are underway to 
identify faint HgMn stars at the VLT so they can thereafter be observed in the
planetary field of CoRot. Whether very-low amplitude modes are detected or not, these 
observations will pose important new constraints on the models.

%

\begin{thebibliography}{99}
\bibitem[1995]{Babel95} Babel, J. 1995, ApJ, 301, 823
\bibitem[2000]{LMR} Leblanc, F., Michaud, G. and Richer, J. 2000, ApJ, 538, 876
\bibitem[1999]{Pamyatnykh95} Pamyatnykh, A. A. 1999, Acta Astronomica, 49, 119
\bibitem[2002]{RMR} Richard, O., Michaud, G. and Richer, J. 2001, ApJ, 558, 377
\bibitem[2000]{RMT} Richer, J., Michaud, G. and Turcotte, S. 2000, ApJ, 529, 338
\end{thebibliography}


\end{document}